\documentclass[runningheads]{llncs}
\usepackage{graphicx}
\usepackage[utf8]{inputenc}
\usepackage{hyperref}
\usepackage{url}
\usepackage{amsmath}
\usepackage{float}
\usepackage{tikz}

\usepackage{listings}
\usepackage[nameinlink]{cleveref}

\usepackage{xspace}

\usepackage{array}
\newcolumntype{L}[1]{>{\raggedright\let\newline\\\arraybackslash\hspace{0pt}}m{#1}}
\newcolumntype{C}[1]{>{\centering\let\newline\\\arraybackslash\hspace{0pt}}m{#1}}
\newcolumntype{R}[1]{>{\raggedleft\let\newline\\\arraybackslash\hspace{0pt}}m{#1}}

\newcommand{\essence}[0]{\textsc{Essence}}
\newcommand{\conjure}{\textsc{Conjure}\xspace}
\newcommand{\savilerow}{\textsc{Savile Row}\xspace}

\newtheorem{exa}{Example}

\usepackage{xcolor}

\begin{document}

\title{Memory Consistency Models using Constraints}
\author{Özgür Akgün \and Ruth Hoffmann \and Susmit Sarkar}

% \titlerunning{Test Case Generation for Memory Consistency Models using Constraints}
% \authorrunning{Özgür Akgün \and Ruth Hoffmann}

\institute{School of Computer Science, University of St Andrews, UK \\
\email{\{ozgur.akgun, rh347, ss265\}@st-andrews.ac.uk}}

\maketitle

\begin{abstract}
Memory consistency models (MCMs) are at the heart of concurrent programming.
They represent the behaviour of concurrent programs at the chip level.
To test these models small program snippets called litmus test are generated, which show allowed or forbidden behaviour of different MCMs.
This paper is showcasing the use of constraint programming to automate the generation and testing of litmus tests for memory consistency models.
We produce a few exemplary case studies for two MCMs, namely Sequential Consistency and Total Store Order.
These studies demonstrate the flexibility of constrains programming in this context and lay foundation to the direct verification of MCMs against the software facing cache coherence protocols.

\end{abstract}

\section{Introduction}

Concurrent programming is orchestrated through the software facing cache coherence protocols and the hardware facing memory consistency models.
The orchestration ensures that the many processors do not interfere with shared resources or receive inconsistent data.
A \emph{memory consistency model} (MCM) describes the observed behaviour on the hardware between  processors and memory.
While a \emph{cache coherence protocol} (CCP) limits the behaviour of the program to mimic the behaviour of the hardware.
Part of the testing of MCMs and CCPs is to check their executions of \emph{litmus tests}.
Litmus tests are small program snippets used to stress particular behaviour of MCMs.
These tests consist of very basic operations between processors and the memory.
Such operations are storing a value to a variable and loading a variable into a register.
Depending on the MCM there are more operations such as fences, but for the scope of this paper we will only consider loads and stores.
Each litmus test has an initial state for all variables and registers.
This can be specified, but in general all variables have the same initial state, with value 0.
Unless otherwise mentioned, we will assume this to be true in the scope of this paper.
For each processor there is a sequence of operations given, which we will call the \emph{program} of the processor.
Each litmus test has a specified final allowed (or prohibited) state of the variables and registers.
A run through a litmus test is called an \emph{execution}.
This final state is found through the concurrent execution of the programs for each processor.
There can be many different final states of the litmus test, as they depend on the concurrent interleaving of the programs from each processor.
Another reason for the variety of final states is the relaxation of the order of execution of the programs within the processors.
These relaxations are defined for each MCM differently, and we will discuss them for our two chosen MCMs in \Cref{sec:sc} and \Cref{sec:tso}.

\begin{exa}
\label{exa:litmus}
\Cref{tab:litmus} shows a litmus test.
This test is over two processors, {\sf{Proc 0}} and {\sf{Proc 1}}.
Each of them is executing a load and a store operation.
The load operation is written as {\sf{MOV EAX $\leftarrow [x]$}} and means that the variable $x$ is read into the register {\sf{EAX}} of a processor.
The store operation is written as {\sf{MOV $[x] \leftarrow$ 2}} and means that the value 2 is stored into the variable $x$.
The variable $x$ is globally visible to all processors.
The final state for this litmus test is given as an allowed state.
We can expect the variable $x$ to contain the value 2, and the register {\sf{EAX}} on {\sf{Proc0}} will contain the value 1, while the register {\sf{EAX}} on {\sf{Proc1}} will contain the value 2.

\begin{table}[H]
\begin{center}
{\textsf{
    \begin{tabular}{|L{3cm}|L{3cm}|}
        \hline
        \multicolumn{1}{|c}{Proc 0} & \multicolumn{1}{c|}{Proc 1} \\ \hline
        MOV $[x] \leftarrow$ 1 & MOV $[x] \leftarrow$ 2  \\ \hline
        MOV EAX $\leftarrow [x]$ & MOV EAX $\leftarrow [x]$ \\ \hline
       \multicolumn{2}{|l|}{Allowed final state} \\
       \multicolumn{2}{|l|}{($x=$ 2 $\land$ Proc0:EAX=1 $\land$ Proc1:EAX=2)} \\ \hline
    \end{tabular}
}}
    \caption{Litmus test example.}
    \label{tab:litmus}
\end{center}
\end{table}
\end{exa}
\vspace{-1.5cm}

Amongst other case studies, this paper will model MCMs as constraint models, and will generate litmus tests which adhere to those models as well as finding all possible final states of variables and registers.

There are currently tools which are approaching the litmus test generation through simulations of MCMs \cite{alglave2014}.
The usage of the toolset provided through {\tt{diy7}} and {\tt{herd7}} is complex and required deep knowledge of the MCMs one wants to work with.
The {\tt{herd7tool}} set is designed to generate and test newly designed MCMs, which are defined in terms of relations between the operations and executions.
The TriCheck tool \cite{trippel2017} verifies the whole system from language to the architecture level, which aids with the evaluation of the compiler.
Both approaches do not use constraint programming, which is our contribution, as the constraint model descriptions for MCMs are short and follow the simple definitions of the MCMs.
The models are written in \conjure{} \cite{akgun2014extensible,akgun2013automated,akgun2011extensible} and translated to input suitable for the constraint solver Minion and the SAT solver lingeling by \savilerow{} \cite{nightingale2017automatically}.

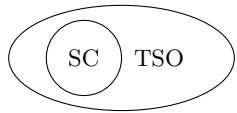
\begin{figure}
    \begin{center}
    \begin{tikzpicture}
        \draw (0.5,0) circle [radius=0.5];
        \draw (1,0) ellipse [x radius=1.5,y radius=0.7];
        \node at (0.5,0) {SC};
        \node at (1.5,0) {TSO};
    \end{tikzpicture}
    \caption{TSO is a relaxation of SC.}
    \label{fig:venn}
    \end{center}
\end{figure}

This paper will contain the \emph{sequential consistency} (SC) model which we discuss in \Cref{sec:sc} and the \emph{total store order} (TSO) model in \Cref{sec:tso}.
TSO is a relaxation of the SC model and thus will allow for more behaviour than SC.
This is reflected in the number of allowed final states from the litmus tests and the number of tests for a given final state.
\Cref{fig:venn} is an illustration of the sets of valid final states of litmus tests depending on the MCM, comparing SC to TSO.
We build \essence{} models of these MCMs, and we will use them to
\begin{itemize}
    \item simulate litmus test executions, for given allowed final state for each MCM
    \item generate litmus tests for given allowed final states, for each MCM
    \item generate all valid executions and all allowed final states, for each MCM
    \item compare litmus tests with allowed behaviour for a MCM against the stricter MCM.
\end{itemize}

These case studies are a stepping stone towards using constraint models to directly and automatically verify MCMs against CCPs.

\section{Memory Consistency Model}
MCMs are describing the observed behaviour of concurrent programs on the hardware chips.
The programs consist of basic load and store operations to and from memory registers.
Depending on the MCM architecture there can be more operations, such as fences, but for our two selected MCMs (SC and TSO) it suffices to only consider loads and stores.
MCMs describe in general the ordering of the operations on a processor level, but also on an execution level.
These two orderings are called \emph{program order} and \emph{memory order} respectively.

\begin{exa}
\label{exa:litmus2}
The litmus test in \Cref{exa:litmus} shows the program order for both {\sf{Proc0}} and {\sf{Proc1}}.
It is the order that the operations are given to the processor, in both cases that is a store followed by a load.
An possible execution of that litmus test is
\begin{table}[H]
\begin{center}
{\sf{
    \begin{tabular}{L{1cm}L{1cm}|R{2.5cm}}
        & Proc0 & MOV $[x] \leftarrow$ 1 \\
        & Proc0 & MOV EAX $\leftarrow [x]$ \\
        & Proc1 & MOV $[x] \leftarrow$ 2 \\
        & Proc1 & MOV EAX $\leftarrow [x]$ \\ \hline
        \multicolumn{3}{c}{\sf{$x=$ 2 $\land$ Proc0:EAX=1 $\land$ Proc1:EAX=2}}.
    \end{tabular}
}}
\end{center}
\end{table}
\vspace{-1cm}

This execution results in the expected final state of {\sf{$x=$ 2 $\land$ Proc0:EAX=1 $\land$ Proc1:EAX=2}}.
Another possible execution of the litmus test is
\begin{table}[H]
\begin{center}
{\sf{
    \begin{tabular}{L{1cm}L{1cm}|R{2.5cm}}
        & Proc0 & MOV $[x] \leftarrow$ 1 \\
        & Proc1 & MOV $[x] \leftarrow$ 2 \\
        & Proc0 & MOV EAX $\leftarrow [x]$ \\
        & Proc1 & MOV EAX $\leftarrow [x]$ \\ \hline
        \multicolumn{3}{c}{\sf{$x=$ 2 $\land$ Proc0:EAX=2 $\land$ Proc1:EAX=2}}.
    \end{tabular}
}}
\end{center}
\end{table}
\vspace{-1cm}

This execution still upholds the program order of both processors, but it violates the expected final state as stipulated in \Cref{tab:litmus}.
\end{exa}

The set of allowed final states of litmus tests is highly dependent on the chosen MCM.
It is possible that we cannot find a valid execution that will lead to the expected allowed final state, within the given restrictions on the ordering of the operations.

\subsection{Sequential Consistency}
\label{sec:sc}
SC is the most restrictive MCM, as it upholds the exact order of operations from the processors to the execution.
Lamport defines in \cite{lamport1979make} \emph{sequential consistency (SC)} to be
``the result of any execution [which] is the same as if the operations of all
the processors were executed in some sequential order, and the
operations of each individual processor appear in this sequence in
the order specified by its program.''

We can look at this in the following way, the order of any two operations has to be kept in the same way as they occur in the litmus test.
It will still allow for the interleaving between the operations of different processors, but not more than that.
This means that both executions as shown in \Cref{exa:litmus2}, are valid SC executions, while having different final states, and thus failing the litmus test in \Cref{tab:litmus}

\begin{exa}
\label{exa:sc}
In addition to the two executions shown in \Cref{exa:litmus2}, below are all executions that can happen in SC when executing the litmus test in \Cref{tab:litmus}.
\begin{table}[H]
{\sf{
    \begin{tabular}{L{1cm}L{1cm}|R{2.5cm}}
        & Proc0 & MOV $[x] \leftarrow$ 1 \\
        & Proc1 & MOV $[x] \leftarrow$ 2 \\
        & Proc1 & MOV EAX $\leftarrow [x]$ \\
        & Proc0 & MOV EAX $\leftarrow [x]$ \\ \hline
        \multicolumn{3}{c}{\sf{$x=$ 2 $\land$ Proc0:EAX=2 $\land$ Proc1:EAX=2}}.
    \end{tabular}
\quad
    \begin{tabular}{L{1cm}L{1cm}|R{2.5cm}}
        & Proc1 & MOV $[x] \leftarrow$ 2 \\
        & Proc1 & MOV EAX $\leftarrow [x]$ \\
        & Proc0 & MOV $[x] \leftarrow$ 1 \\
        & Proc0 & MOV EAX $\leftarrow [x]$ \\ \hline
        \multicolumn{3}{c}{\sf{$x=$ 1 $\land$ Proc0:EAX=1 $\land$ Proc1:EAX=2}}.
    \end{tabular}

}}
\end{table}
\vspace{-1cm}
\begin{table}[H]
{\sf{
    \begin{tabular}{L{1cm}L{1cm}|R{2.5cm}}
        & Proc1 & MOV $[x] \leftarrow$ 2 \\
        & Proc0 & MOV $[x] \leftarrow$ 1 \\
        & Proc1 & MOV EAX $\leftarrow [x]$ \\
        & Proc0 & MOV EAX $\leftarrow [x]$ \\ \hline
        \multicolumn{3}{c}{\sf{$x=$ 1 $\land$ Proc0:EAX=1 $\land$ Proc1:EAX=1}}.
    \end{tabular}
\quad
    \begin{tabular}{L{1cm}L{1cm}|R{2.5cm}}
        & Proc1 & MOV $[x] \leftarrow$ 2 \\
        & Proc0 & MOV $[x] \leftarrow$ 1 \\
        & Proc0 & MOV EAX $\leftarrow [x]$ \\
        & Proc1 & MOV EAX $\leftarrow [x]$ \\ \hline
        \multicolumn{3}{c}{\sf{$x=$ 1 $\land$ Proc0:EAX=1 $\land$ Proc1:EAX=2}}.
    \end{tabular}
}}
\end{table}
\vspace{-1cm}
\end{exa}

\subsection{Total Store Order}
\label{sec:tso}
As SC is very strict it is not observed on manufactured chip sets.
A first relaxation of the MCM is to allow a reordering of the store and load order.
This relaxation has been first observed on the Intel Sparc architecture and is called the \emph{Total Store Order (TSO)} \cite{sparc1994}.
Similarly to SC it restricts three of the four possible orderings of loads and stores to be the same in program order as in memory order but the Store $\rightarrow$ Load program order can be disregarded in the memory order.
This relaxation allows for more possible executions and more allowed final states of litmus tests.

\begin{exa}
\label{exa:tso}
As TSO is a relaxation of SC, it will still observe the same executions of the litmus test in \Cref{exa:litmus}, as listed in \Cref{exa:litmus2} and \Cref{exa:sc}.
In addition it can observe the following execution (amongst others)
\begin{table}[H]
\begin{center}
{\sf{
    \begin{tabular}{L{1cm}L{1cm}|R{2.5cm}}
        & Proc0 & MOV $[x] \leftarrow$ 1 \\
        & Proc0 & MOV EAX $\leftarrow [x]$ \\
        & Proc1 & MOV EAX $\leftarrow [x]$ \\
        & Proc1 & MOV $[x] \leftarrow$ 2 \\ \hline
        \multicolumn{3}{c}{\sf{$x=$ 2 $\land$ Proc0:EAX=1 $\land$ Proc1:EAX=1}}.
    \end{tabular}
}}
\end{center}
\end{table}
\vspace{-1cm}

It is impossible to find a valid SC execution of the litmus test in \Cref{exa:litmus} that will result in this final state.
\end{exa}

We do note that this simple, yet formal description of TSO is still stricter than what is observed on some TSO architecture (for example x86).
As we do not implement fences in our current constraint model of TSO we will stick to the textbook definition of TSO as given above.

\section{A Generic MCM Model}

\Cref{fig:essence1} shows the initial setup of the various components needed for any MCM constraint model.
Each constraint model that we create for any MCM is defined over a finite number of cores (or processors), registers per core, variables and a finite number of values for the variables.
We also have to define the maximal number of operations that we can have in a litmus test, per processor.
As mentioned above the only operations that we consider are stores and loads.
Each store assigns a value to a global variable, and each load loads the variable into a processor's register.
Depending on the goal of the model, we will either generate a program (litmus test), or find an execution of a litmus test.
Each program consists of a sequence of operations for each core.
Whereas an execution is a sequence of all operations in all cores combined.
The maximal size of an execution of a litmus test will be the number of processors times the maximal number of operations per processors.
Note that we use core and processor, and program and litmus test synonymously here.

A program is represented as a sequence of operations per core. An operation may either be a load or a store, so we use a \emph{variant type} to denote these. A variant type models a tagged union, in this case load and store are the tags. For a decision variable \texttt{x} with the domain \texttt{OPERATION}, we determine whether a given tag is active using the \texttt{active} operator of \essence{}: \texttt{active(x, load)} is true if and only if \texttt{x} denotes a load operation. When modelling with variant types we tend to use the \texttt{active} operator often together with a logical implication, where the left-hand-side of the implication determines which tag is active and the right-hand-side posts a constraint assuming that particular tag to be active. Syntactical support for a switch-case style discriminator for variant types does not exist in \essence{} yet.

\begin{figure}
\begin{lstlisting}
language Essence 1.3

given CORE new type enum
given REGISTER new type enum
given VARIABLE new type enum
given VALUE new type enum
given maxNbOperationsPerCore : int(1..)

letting MCMs be new type enum {SC, TSO}
given MCM : MCMs

letting OPERATION be domain
 variant
  $ load the value stored in the variable into the register
  { load  : record { register : REGISTER
                   , variable : VARIABLE }

  $ store the value into the variable
  , store : record { variable : VARIABLE
                   , value    : VALUE }
  }

find program :
 function (total) CORE -->
  sequence (minSize 1, maxSize maxNbOperationsPerCore)
   of OPERATION

letting maxNbOperationsInExec be
 1 + maxNbOperationsPerCore * sum([1 | c : CORE])

find execution :
 sequence (maxSize maxNbOperationsInExec) of
  (CORE, OPERATION)
\end{lstlisting}
\caption{Initial setup of variables and programs for any MCM.\label{fig:essence1}}
\end{figure}

In \Cref{fig:essence2} we constrain that every operation in the litmus test can only occur exactly once in the execution.
Each operation in a litmus test is unique as it has an implicit index.
So should there be for example two distinct load operations of the same variable into the same register on a core in a litmus test, then those two load operations are treated as distinct in the execution.
In addition, there is the implied constraint which states that when every operation in the program is occurring exactly once in an execution then the length of the execution is exactly the number of all operations of all cores in the program.

\begin{figure}
\begin{lstlisting}
$ every operation in program has to exist once in execution
such that
 forAll (core, operations) in program .
  forAll (index, operation) in operations .
   1 = (sum (indexE, (coreE, operationE)) in execution .
     toInt(core = coreE /\ operation = operationE))

$ the length of execution in terms of the length of program
such that
 |execution| = sum([ |ops| | (c,ops) <- program ])
\end{lstlisting}
\caption{Restrictions on execution of litmus tests.\label{fig:essence2}}
\end{figure}

As the execution progresses we need to keep track of the registers and variables.
Each operation might alter the state of them.
We also need to note that both variables and registers start off in an independent initial state, which is why in \Cref{fig:essence4} the length of the sequences of all states of variables and registers is one longer than the execution, as the initial state exists before the execution starts.

\begin{figure}
\begin{lstlisting}
find state_of_registers :
 sequence (maxSize maxNbOperationsInExec) of
  function (total) (CORE, REGISTER) --> VALUE

find state_of_variables :
 sequence (maxSize maxNbOperationsInExec) of
  function (total) VARIABLE --> VALUE

such that
 |execution| + 1 = |state_of_variables|,
 |execution| + 1 = |state_of_registers|
\end{lstlisting}
\caption{Registers and variables setup. \label{fig:essence4}}
\end{figure}

The initialisation of the variables and registers is shown in \Cref{fig:essence-step1}.
The registers have a special empty/initial value given to them, which is different to the value a variable can contain at anytime, including the initial value a variable might contain.
The initial value of a variable might be reassigned to the variables in the course of a program, whereas the initial value of the register cannot be re-attained.

\begin{figure}[t]
\begin{lstlisting}
$ defining the state_of_registers and state_of_variables in terms of the execution
$ define step(1) as the initial step
such that
 $ initially, all registers are empty
 forAll r : REGISTER . forAll c : CORE .
  state_of_registers(1)((c,r)) = initial_state_of_registers,

 $ initially, all variables are 0
 forAll v : VARIABLE .
  state_of_variables(1)(v) = initial_state_of_variables,

 $ variables cannot be assigned to initial_state_of_registers
 forAll (index, mapping) in state_of_variables .
  forAll (var, val) in mapping . val != initial_state_of_registers,

 $ registers cannot be assigned to initial_state_of_registers after step 1
 forAll (index, mapping) in state_of_registers .
  index != 1 ->
  forAll ((core, reg), val) in mapping .
   $ if it takes the INITIAL value now, it must have been INITIAL always
   initial_state_of_registers = val ->
   initial_state_of_registers =
       state_of_registers(index-1)((core, reg))
\end{lstlisting}
\caption{Variable and register initialisations.\label{fig:essence-step1}}
\end{figure}

\Cref{fig:essence-step2} describes what happens to the variables and registers when an operation occurs.
If the operation is a load, there is no change to the variables.
Only the register on the core that the load has been called from changes.
The next state of that register will contain the variable loaded.
If the operation is a store, there is not change to any of the registers.
Only the variable that the store occurs on will change.
The next state of that variable will contain the value assigned to it in the store.

\begin{figure}
\begin{lstlisting}
$ defining the state_of_registers and state_of_variables
$ in terms of the execution
$ define step(index+1) in terms of step(index)
$ and execution(index)
such that
 forAll (index, (core, operation)) in execution .
  and([
   active(operation, load) ->
    and([ state_of_registers(index+1)((core, operation[load][register])) =
       state_of_variables(index)(operation[load][variable])
     $ all other registers on this core stay the same
     , forAll r : REGISTER .
      r != operation[load][register] ->
      state_of_registers(index+1)((core, r)) = state_of_registers(index)((core, r))
     $ all variables stay the same
     , forAll v : VARIABLE .
      state_of_variables(index+1)(v) = state_of_variables(index)(v)
     ])
   , active(operation, store) ->
    and([ state_of_variables(index+1)(operation[store][variable]) =
       operation[store][value]
     $ all registers on this core stay the same
     , forAll r : REGISTER .
      state_of_registers(index+1)((core, r)) = state_of_registers(index)((core, r))
     $ all other variables stay the same
     , forAll v : VARIABLE .
      v != operation[store][variable] ->
      state_of_variables(index+1)(v) = state_of_variables(index)(v)
     ])
   $ all registers on other cores stay the same
   , forAll c : CORE . c != core ->
    forAll r : REGISTER .
     state_of_registers(index+1)((c, r)) = state_of_registers(index)((c, r))
   ])
\end{lstlisting}
\caption{Variable and register updates. \label{fig:essence-step2}}
\end{figure}

As we are interested in the different allowed final states of a program we include parameters limiting the set of allowed final states.
In \Cref{fig:essence6} we define the constraints that we have to reach a given final state.
We also allow for the initial states of the variables and registers to be defined in a flexible way, but for the set of tests in \Cref{sec:appl} we kept the initial states to be the same.
The include and exclude program sets are used to compare the different MCMs to each other.
We will generate a larger set of litmus tests that all hold for the relaxed TSO and filter out the tests that do not hold in the stricter SC model.

\begin{figure}[t]
\begin{lstlisting}
$ a way to post constraints on the initial and the final
$ values that are stored in each register and variable
$ doesn't have to mention all registers.
given initial_state_of_registers, initial_state_of_variables : VALUE
given final_state_of_registers : function (CORE, REGISTER) --> VALUE
such that
 forAll ((c,r),v) in final_state_of_registers . state_of_registers(|state_of_registers|)((c,r)) = v
given final_state_of_variables : function VARIABLE --> VALUE
such that
 forAll (var, val) in final_state_of_variables . state_of_variables(|state_of_variables|)(var) = val


given include_programs : set of function (total) CORE --> sequence of OPERATION
such that |include_programs| > 0 -> program in include_programs

given exclude_programs : set of function (total) CORE --> sequence of OPERATION
such that !(program in exclude_programs)

branching on [program]
\end{lstlisting}
\caption{Final allowed state.\label{fig:essence6}}
\end{figure}

\subsection{Sequential Consistency}
As introduced in \Cref{sec:sc} SC defines that any two operations of a core will occur in an execution in the same order as they are written in the program for each core.
In \Cref{fig:essenceSC} we define the constraints which enforces SC in the execution of a litmus test.
We use the index of the execution, which is a sequence of operations, to track the ordering of the operations in relation to their index in the program on the core.
Operations from other cores can be interleaved.

\begin{figure}
\begin{lstlisting}
$ SC: Sequential Consistency
$ Every pair of operation from the program will have the same order in execution
such that MCM = SC ->
 forAll (core, operations) in program .
  forAll index1, index2 in defined(operations) .
   index1 < index2 ->
    exists indexE1, indexE2 in defined(execution) .
     indexE1 < indexE2 /\
     execution(indexE1) = (core, operations(index1)) /\
     execution(indexE2) = (core, operations(index2))
\end{lstlisting}
\caption{Constraints defining sequential consistency in the execution. \label{fig:essenceSC}}
\end{figure}

\subsection{Total Store Order}
In \Cref{sec:tso} we discussed TSO, which is a relaxation of SC.
\Cref{fig:essenceTSO} uses the index of the operation in the program against the index in the execution sequence.
In addition it uses the active constraint to check whether we can relax that ordering for a store load pair of operations.

\begin{figure}
\begin{lstlisting}
$ TSO: Total Store Order
such that MCM = TSO ->
 forAll (core, operations) in program .
  forAll index1, index2 in defined(operations) .
   index1 < index2 /\
   $ if the first is a store, and the second is a load, do not post any constraints
   !(active(operations(index1), store) /\
       active(operations(index2), load)) ->
    exists indexE1, indexE2 in defined(execution) .
     indexE1 < indexE2 /\
     execution(indexE1) = (core, operations(index1)) /\
     execution(indexE2) = (core, operations(index2))
\end{lstlisting}
\caption{Constraints defining total store order in the execution.\label{fig:essenceTSO}}
\end{figure}

\section{Case Studies}
\label{sec:appl}
All the files generated and experiments mentioned can be found in a github repository\footnote{\url{https://github.com/stacs-cp/ModRef2018-MCM}} and can be reproduced using \conjure{}, \essence{}, \savilerow{}, lingeling and the {\tt{herd7tool}} sets.

\subsection{Litmus Test Simulation}
We used the {\tt{herd7tool}} set \cite{alglave2014} to generate litmus tests for TSO and SC.
\Cref{fig:herd} shows the commands used for the litmus test generation using {\tt{diy7}}.
The litmus tests generated for TSO included tests with fences, which we ignored.
This resulted in 65 litmus tests.
\begin{figure}
\begin{lstlisting}
# SC litmus test generation
> diy7 -arch X86 -nprocs 4 -size 6 -mode uni

# TSO litmus test generation
> diy7 -arch X86 -nprocs 4 -size 6 -safe Rfe,Fre,Wse,PodR*,PodWW,MFencedWR -relax PodWR,[Rfi,PodRR]
\end{lstlisting}
\caption{{\tt{diy7}} litmus test generation commands.\label{fig:herd}}
\end{figure}

We translated the litmus tests into a \essence{} parameter file.
The litmus test in \Cref{exa:litmus} is the generated {\tt{SB000a}} litmus test.
The corresponding \essence{} parameter file is shown in \Cref{fig:essenceparam}.
We set the cores, registers, variables and available values, and we limit the maximal number of operations per core.
The initial states of both the registers and the variables are also given.
We take the final states of the registers and variables from the generated litmus test.
We can choose which MCM we want to check against.
In {\tt{include\_programs}} we have the litmus test, which consists of the sequence of operations per core.
The {\tt{exclude\_programs}} set is empty, as it is only used when checking a larger set of litmus tests that can be generated and are valid in one MCM against a more restrictive MCM.

\begin{figure}
\begin{lstlisting}
letting CORE be new type enum {c1, c2}
letting REGISTER be new type enum {r1, r2}
letting VARIABLE be new type enum {x}
letting VALUE be new type enum {Initial, v0, v1, v2}

letting maxNbOperationsPerCore be 2

letting initial_step_of_registers be Initial
letting initial_step_of_variables be v0
letting final_step_of_variables be function
 ( x --> v2 )
letting final_step_of_registers be function
 ( (c1,r1) --> v1
 , (c2,r2) --> v2
 )

letting MCM be SC
$ letting MCM be TSO

letting include_programs be {
 function(c1 --> sequence( variant {store = record {variable = x, value = v1}}
    , variant {load = record {register = r1, variable = x}}
 )
 ,c2 --> sequence( variant {store = record {variable = x, value = v2}}
    , variant {load = record {register = r2, variable = x}}
 ))
}

letting exclude_programs be {}
\end{lstlisting}
\caption{\essence{} parameter file of a litmus test.\label{fig:essenceparam}}
\end{figure}

We used these 65 tests to simulate and check our implementation of the SC and TSO.
We have found that our constraint model of SC (\Cref{fig:essenceSC}) is valid on the same 29 tests as the {\tt{herd7}} SC model described in terms of the relations between operations in \Cref{fig:herdSC}.

\begin{figure}
\begin{lstlisting}
include "fences.cat"
include "cos.cat"

(* Atomic *)
empty rmw & (fre;coe) as atom

(* Sequential consistency *)
acyclic po | fr | rf | co as sc
\end{lstlisting}
\caption{{\tt{herd7}} model of SC.\label{fig:herdSC}}
\end{figure}

Similarly, our TSO model in \Cref{fig:essenceTSO} corresponds to a more restricted version of the x86-TSO model, as we currently still restrict the ordering of stores followed by loads from external cores.
This is apparent by running the {\tt{herd7}} TSO model in \Cref{fig:herdTSO} which accepts the same 51 tests as our \essence{} TSO model.
\begin{figure}
\begin{lstlisting}
include "cos.cat"

(* Communication relations that order events*)
let com-tso = rf | co | fr
(* Program order that orders events *)
let po-tso = po & (W*W | R*M)

(* TSO global-happens-before *)
let ghb = po-tso | com-tso
acyclic ghb as tso
show ghb
\end{lstlisting}
\caption{Strict TSO model in {\tt{herd7}}.\label{fig:herdTSO}}
\end{figure}

\subsection{Litmus Test Generation}
Having shown that our SC and TSO models are correctly reflecting the behaviour, we have gone further to generate more litmus test using constraint solvers.
Having a greater number of litmus tests allows for a greater coverage of possible behaviour exhibited by an MCM.
We produced sets of litmus tests which were given a static number of cores, registers, variables, values and a maximal number of operations per core.
In addition, we specified the allowed final state of the litmus test.
We let the solvers generate a set of litmus tests that had at least one valid execution in TSO with the allowed final state.
Then we filtered those tests to see which would still have a valid execution in SC.
We have done this for two different scenarios.
Both have 2 cores, 2 registers per core, 2 variables and 2 possible values.

In the first scenario each core can have up to 2 operations, and we have set the final state of the registers to be that one register on one core contains an initial value and one register of the other core contains the value 1.
For TSO this resulted in a total of 160 litmus test, of which 132 have a valid SC execution.

The second scenario has for each core up to 3 operations and the final state for the registers is restricted to the first register of both cores having the value 0 and the second register on both cores to have the value 1.
We generated 1154 litmus test with a valid execution in TSO, and thereof 776 have a valid execution in SC.

\subsection{Output Generation}
Any litmus test consists of a program per core and an allowed final state of any execution.
We pursued to see how many valid and distinct final states can be found for a given program, depending on the MCM.
We used the generated litmus tests from {\tt{diy7}} and ignored the required final states.
The solvers now use the models for TSO and SC to find all executions with distinct allowed final states, which are a combination of both variables and registers and do not have to be enforced on all variables and registers available.

We find that for all litmus tests there is always a valid execution with a valid final state, for both SC and TSO.

The highest number of distinct final states for a litmus test is 72 ({\tt{IRIW000}}).
This number of distinct final states is true for both TSO and SC.
The litmus test with the next highest number of possible final states (63) for TSO is {\tt{3.SB000}}, which only has 7 distinct final states for SC.

For SC most litmus tests have less than 10 distinct final states.
While this observation can be also made for TSO, there are fewer litmus tests which have only 1 distinct final state.

\section{Conclusion}
We have shown that using constraint programming to model MCMs and to use them to generate litmus tests is a viable alternative to algorithms and other tools.
Constraint programming is a highly flexible way to re-use the models for different scenarios and different uses.
We also find that the models created are a lot smaller than what can be created with model checking. As in that context the models face the state explosion problem.
Our aim with this work is to be able to implement cache coherence protocols along side the MCMs and to directly verify these two models against each other.
Work in this direction but using model checking and a few additional manual proofs were done in \cite{banks2017verification}.
We believe constraint programming to be an ideal contender for this problem, as our current models suggest to be smaller than the model checking models.

\bibliographystyle{splncs04}
\bibliography{mcm}

\end{document}